\documentclass[a4paper]{llncs}
\usepackage{amssymb,amsmath,amsfonts}
\usepackage{comment}

\newcommand{\set}[2]{\left\{#1\ \left|\ \vphantom{#1} #2\right.\right\}}
\newcommand{\N}{\ensuremath{\mathbb N}}
\newcommand{\Z}{\ensuremath{\mathbb Z}}

\usepackage{ifpdf}
\ifpdf
  \usepackage[pdftex]{graphicx}
  \usepackage[pdftex,colorlinks,bookmarksopen,pdfpagemode=UseOutlines]{hyperref}
  \hypersetup{pdfnewwindow=true}
\else
  \usepackage{psfig}\pssilent
  \usepackage[dvips]{hyperref}
\fi

\usepackage{xspace}

\newcommand{\Ttr}{T_{\text{tr}}}
\newcommand{\Tech}{T_{\text{samp}}}
\newcommand{\reglenum}[1]{\ensuremath{\mathfrak{#1}}}
\newcommand{\rn}{\reglenum}
\newcommand{\rnnl}{$\rn{9}_\ell$\xspace}  
\newcommand{\rnnh}{$\rn{9}_h$\xspace}
\newcommand{\rncdl}{$\rn{110}_\ell$\xspace}  
\newcommand{\rncdh}{$\rn{110}_h$\xspace}
\newcommand{\rncvsl}{$\rn{126}_\ell$\xspace}  
\newcommand{\rncvsh}{$\rn{126}_h$\xspace}

\newcounter{ifnote}
\usepackage{ifthen}
\usepackage{calc}	    	

\reversemarginpar \marginparwidth 3.1cm
\newcounter{Note}[page]  \newcounter{notesimple}
\newcounter{notemark}  \newcounter{notemarkref}
 \newcounter{notetext} 
\newcounter{notetextref} 

\newcommand{\notemark}{%
	\ifthenelse{\equal{\value{notemark}}{\value{notemarkref}}}{%
    \refstepcounter{notemark}%
    \refstepcounter{notemarkref}%
    \setcounter{Note}{\value{notemark}}%
  }{%
    \refstepcounter{notemark}%
    \setcounter{Note}{\value{notemark}}%
	}%
	\setcounter{notetextref}{\value{notemarkref}}%
	\begin{picture}(0,0)%
    \put(-3,-3){\LARGE\theNote}
	\end{picture}\xspace}

\newcommand{\notetextbase}[1]{%
  \ifthenelse{\value{ifnote} = 1}{\marginpar{\notedebasenumero{#1}}}{}}

\newcommand{\notetext}[1]{
  \ifthenelse{\equal{\value{notetext}}{\value{notetextref}}}{%
    \refstepcounter{notetext}%
    \refstepcounter{notetextref}%
    \setcounter{Note}{\value{notetext}}
  }{
    \refstepcounter{notetext}%
    \setcounter{Note}{\value{notetext}}}%
  \setcounter{notemarkref}{\value{notetextref}}%
  \notetextbase{#1}}

\newcommand{\notedebase}[1]{%
	\raggedright%
  \footnotesize%
  #1 \\%
  \rule[1.4mm]{\linewidth}{0.5pt} \\%
}

\newcommand{\notedebasenumero}[1]{%
  \theNote{} \notedebase{#1}}

\newcommand{\notesimple}[1]{%
	\ifthenelse{\value{ifnote}=1}{\marginpar{\notedebase{#1}}}{}}

\newcommand{\notenumero}[1]{%
  \ifthenelse{\value{ifnote} = 1}{%
    \refstepcounter{Note}%
    \setcounter{notemarkref}{\value{Note}}%
    \setcounter{notetextref}{\value{Note}}%
    \setcounter{notemark}{\value{Note}}%
    \setcounter{notetext}{\value{Note}}%
    \begin{picture}(0,0)%
      \put(-3,-3){\LARGE\theNote}
    \end{picture}%
    \ifthenelse{\isodd{\value{Note}}}{%
      \protect\reversemarginpar%
      \marginpar[{\notedebasenumero{#1}}]{\notedebasenumero{#1}}}{%
      \protect\normalmarginpar%
      \marginpar[{\notedebasenumero{#1}}]{\notedebasenumero{#1}}}}{}}

\newcommand{\todo}[2][1]{%
	\ifthenelse{\equal{#1}{0} \or \equal{#1}{simple}}{%
    \setcounter{notesimple}{1}}{}%
	\ifthenelse{\value{notesimple}=0}{\notenumero{#2}}{\notesimple{#2}}}

\usepackage{picins}

\begin{document}
\pagestyle{empty}
\mainmatter
\title{Coalescing Cellular Automata}

\authorrunning{Jean-Baptiste Rouquier and Michel Morvan}
\author{Jean-Baptiste Rouquier\inst{1} and Michel Morvan\inst{1,2}}

\institute{ENS Lyon, LIP,
46 all\'ee d'Italie,
69364 Lyon, France
\and
EHESS and Santa Fe Institute.
\\ \email{\{jean-baptiste.rouquier, michel.morvan\}@ens-lyon.fr}}

\maketitle
\begin{abstract}

We say that a Cellular Automata (CA) is coalescing when its execution on two
distinct (random) initial configurations in the same asynchronous mode
(the same cells are updated in each configuration at each time step)
makes both configurations become identical after a reasonable time.
We prove coalescence for two elementary rules and show
that there exists infinitely many coalescing CA.
We then conduct an experimental study on all elementary CA and show that some rules exhibit a phase
transition, which belongs to the universality class of directed percolation.

\end{abstract}

\section{Introduction}
The \emph{coalescence} phenomenon, as we call it, has been observed for the
first time by Nazim Fates \cite{async_eca_density_robust}, in the context of
asynchronous cellular automata. Coalescing CA exhibit the
following behavior: starting from two different initial random configurations
and running the same updating sequence
(the same cells are updated at each time step in both configurations),
the configurations quickly become identical, i.e. the
dynamics not only reach the same attractor, they also synchronize their
orbits. This of course appears in trivial situations, for example if the CA
converges on a single fixed point, but Nazim Fates has also observed it in a case
where the coalescing orbit is absolutely non trivial.

The goal of this paper is to explore this rather strange emergent phenomenon in
which the asymptotic behavior seems to be only related to the (random) sequence of
update of the cells and not to the initial configuration.
This work shows that, in some cases, the randomness used during evolution is as important as the
one used during initialization: this stochastic dynamic, with high entropy, is
perfectly insensitive to initial condition (no chaos here).

The results presented here are of two kinds. First, we prove the existence of
infinitely many different (we precise this notion) non trivial coalescent
CA. Secondly, we study by simulation the behavior of all elementary CA (ECA) with
regards of this coalescence property, in an asynchronous context in which at each
step each cell has a fixed probability $\alpha$ to be updated. We show that over
the 88 different ECA, six situations occur: a/ 37 ECA never coalesce; b/ 20 always
coalesce in a trivial way (they converge to a unique fixed point); c/ 6 always
coalesce on non trivial orbits; d/ 14 combine a/, b/ and c/ depending on
$\alpha$ (4 combine a/ and b/, 3 combine b/ and c/, and 7 combine a/ and c/); e/
7 enter either full agreement (coalescence) or full disagreement; last, f/
4 combine e/ with either a/, b/ or c/.

We also study the transition between non coalescence and
coalescence when $\alpha$ varies for the ECA that combine a/ and c/:
there is
a phase transition belonging to the universality class of
directed percolation.
We thus get a new model of
this class,
with a few variants.
An unusual fact among directed percolation models is that the limit of the
sub-critical regime is neither a single absorbing state, nor a set of fixed
points, but a non trivially evolving phase.
Its originality and links to other domains could help understanding this
class and hopefully
lead to analytical results.

\medskip
The paper is organized as follows. Section~\ref{defs} gives definitions and
notations.
We prove in section~\ref{formal-proof} that, under certain conditions, CA \rn6
and \rn7 (using the Wolfram's numbering of ECA) are coalescing and show how to
construct from them coalescing CA with arbitrarily many states.
We also prove
that CA \rn{15} and \rn{170} either coalesce or enter total
disagreement, each case occurring with probability $\frac12$.
In Section~\ref{expérimental-study}, we describe the exhaustive simulation study
of all ECA and then check the directed percolation hypothesis.
Moreover, we prove that some CA exhibit two phase
transitions: one for small $\alpha$ and one for high $\alpha$.

\section{Definitions and Notations}
\label{defs}
In this paper, we consider the dynamics of some CA when they are run on an
asynchronous mode. Let us start by defining the synchronisms we work with.

\begin{definition}
  An
  \emph{asynchronous finite CA} is
  a tuple $(Q,\,d,\,V,\,\delta,\,n,\,\mu)$ where
  \begin{itemize}
  \item $Q$ is the set of \emph{states};
  \item $d\in\N^*$ is the \emph{dimension};
  \item 
    $V=\{v_1,\,\dots,\,v_{|V|}\}$, the \emph{neighborhood},
    is a finite set of vectors in $\Z^d$;
  \item $\delta:Q^{|V|}\to Q$ is the \emph{transition rule};
  \item $n\in\N^*$ is the \emph{size};
  \item ${\cal U}:=(\Z/n\Z)^d$ is the \emph{cell space} (with periodic boundary condition);
  \item $\mu$, the \emph{synchronism}, is a probability measure on $\{0,1\}^{\cal U}$.
  \end{itemize}
A \emph{configuration} specifies the state of each cell,
and so is a function
$c:{\cal U}\to Q$.
\end{definition}

The dynamic is then the following. Let $c_t$ denote the configuration at time $t$,
($c_0$ is the initial configuration).
Let $\set{M_t}{t\in\N}$ be a sequence of independent identically distributed random variables with
distribution $\mu$.
 The configuration at time $t+1$ is obtained by
\begin{equation*}
  c_{t+1}(z):=
  \begin{cases}
    c_t(z) & \text{if } M_t(z)=0\\
    \delta\big(c_t(z+v_1),\,\dots,\,c_t(z+v_{|V|})\big) & \text{if } M_t(z)=1\\
  \end{cases}\ .
\end{equation*}
In other words, for each cell $z$, we apply the usual transition rule if
$M_t(z)=1$ and freeze it (keep its state) if $M_t(z)=0$.

\medskip
Here are
the two synchronisms we use.
They are the most natural, even if others are possible
(like systematic or alternating sweep, or updating some cells more often,
but this require non independent $M_t$ and so a more general formalism).
%
If $x\in\{0,1\}^{\cal U}$, let $|x|_1$ be the number of 1 in the coordinates of $x$.

\emph{The Partially Asynchronous Dynamic}
Let $0<\alpha\le 1$.
For each cell, we update it with probability $\alpha$, independently from its
neighbors.
$\mu$ is thus the product measure of Bernoulli distributions:
$\mu(x):=\alpha^{|x|_1}(1-\alpha)^{|x|_0}$ (with $0^0=1$).
The case $\alpha=1$ corresponds to the synchronous dynamic.

\emph{The Fully Asynchronous Dynamic}
At each step, we choose one cell and update it. Which defines
$\mu(x)$ as $1/n$ if $|x|_1=1$ and $\mu(x):=0$ otherwise.

\medskip
We now introduce the definition of coalescing CA to formalize the observation of~\cite{async_eca_density_robust}.
The principle is to use two initial configurations, and to let them evolve with the
\emph{same} outcome of the random variables $\set{M_t}{t\in\N}$.
In other words, we use two copies of the CA, and at each time step,
we update
the same cells in both copies.
This comes down to using the same source of randomness for both copies, like
in~\cite{sync-by-id-noise} (on another system) where the authors observe a synchronization.
\begin{definition}
  An asynchronous finite CA is \emph{coalescing} if, for any two initial 
  configurations,
  applying the same sequence of updates leads both configurations
  to become identical within
  polynomial expected time (with respect to $n$).
\end{definition}
Any nilpotent CA (converging toward a configuration where all states are
identical) is coalescing if it converges in polynomial time. But there are non nilpotent coalescing CA,
which we call non trivial.
We now consider only those CA.


In the following, we heavily use the simplest CA, namely the Elementary CA:
one dimension ($d=1$), $2$ states ($Q=\{0,1\}$), nearest neighbors ($V=\{-1,0,1\}$).
There are $2^8=256$ possible rules, $88$ after symmetry
considerations.
We use the notation introduced by S. Wolfram, numbering the rules from \rn0 to
\rn{255}.

\section{Formal Proof of Coalescence}
\label{formal-proof}
In this section, we prove that there are infinitely many coalescing CA. For
that, we prove the coalescence of two particular CA and show how to build an
infinite number of coalescing CA from one of them.
An easy way to do that last
point would be to extend a coalescing CA by adding states that are always mapped
to one states of the original CA,
regardless of their neighbors.
However, we consider such a transformation to be artificial since it
leads to a CA that is in some sense identical.
To avoid this,
we focus on
\emph{state minimal} CA: CA in which any state can be reached.
Note that among ECA, only \rn0 and \rn{255} are not state-minimal.

We
first exhibit two state-minimal coalescing CA
(proposition~\ref{6-et-7-coalescent}); then, using this result, deduce the
existence of an infinite number of such CA (theorem~\ref{main-theo}); and
finally describe the coalescent behavior of two others ECA
(proposition~\ref{15-170-semi-coalescent}).
\begin{proposition}
  \label{6-et-7-coalescent}
  Rules \rn{6} and \rn{7} are coalescing for the fully asynchronous dynamic
  when $n$ is odd.
\end{proposition}

\begin{proof}
We call \emph{number of zones} the number of patterns $01$ in a configuration,
which is the number of ``blocks'' of consecutive $1$ (those blocks are the
zones). We first consider only one copy (one configuration).

\smallskip
\parpic[r]{
\begin{tabular}{c||c|c|c|c|c|c|c|c}
Neighbors &1\,1\,1&1\,1\,0&1\,0\,1&1\,0\,0&0\,1\,1&0\,1\,0&0\,0\,1&0\,0\,0\\
  \hline
  New state &  0  &  0  &  0  &  0  &  0  &  1  &  1  &  0\\
\end{tabular}}
Here is the transition table of \rn{6}.
Since one cell at a time is updated, and updating the central cell of $101$
or $010$ does not change its state, zones cannot merge.

\parpic[r]{$
  \begin{matrix}
    \cdots\ 0\ \underbrace{0\ 0\ 1}\ \cdots\\
\cdots\ \underbrace{0\ 0\ 1}\ 1\ \cdots\\
\cdots\ 0\ \underbrace{1\ 1\ 1}\ \cdots\\
\cdots\ 0\ 1\ 0\ 1\ \cdots
\end{matrix}
$}
On each pattern $111$, the central cell can be updated (leading to the pattern $101$)
before its neighbors (with probability $\frac13$) with expected time $n$.
On each pattern $0001$, the opposite sequence is possible.
It happens without other update of the four cells with probability $1/4^3$ and
with an expected time of $3\,n$.
So, as long as there are patterns $000$ or $111$, the number of zones increases
with an expected time $O(n)$. Since there are $O(n)$ zones, the total expected
time of this increasing phase is $O(n^2)$.

\medskip
The configuration is then regarded as a concatenation of words on $\{0,\,1\}^*$.
Separation between words are chosen to be the middle of each pattern $00$ and $11$,
so we get a sequence of
words that have no consecutive identical letters, each word being at least two
letter long (that is, words of the language ``$(01)^+0?\ |\ (10)^+1?$'').
We now show that borders between these words follow a one way random walk
(towards right) and
meet, in which case a word disappear with positive probability.
The CA evolves therefore towards a configuration with only one word.

Updating the central cell of $100$ does not change its state, so the borders
cannot move towards left more than one cell.
On the other hand, updating the central cell of $001$ or $110$ make the border move.
One step of this random walk takes an expected time $O(n)$.

The length of a word also follows a (non-biased) random walk, which reaches $1$
after (on average) $O(n^3)$ steps, leading to the pattern $000$ or $111$.
This pattern disappears with a constant non zero probability like in the
increasing phase.
The expected time for $O(n)$ words to disappear is then $O(n^4)$.

\medskip
Since $n$ is odd, the two letters at the ends of the words are the same, i.e.
there is one single pattern $00$ or $11$, still following the biased random walk.
We now consider again the two copies.
This pattern changes the phase in the sequence $(01)^+$, it is therefore a
frontier between a region where both configurations agree and a region where
they do not.
The pattern in the other configuration let us come back to the region where the
configurations have coalesced.
We study the length of the (single) region of disagreement.
It follows a non biased random walk determined by the moves of both patterns.

\parpic[r]{$
  \begin{matrix}
    \cdots\ 1\ 0\ 1\ 0\ 0\ 1\ 0\ 1\ \cdots\\[-0.5ex]
    \cdots\ 0\ 1\ 0\ 1\ 1\ 0\ 1\ 0\ \cdots
  \end{matrix}
  $}
When this length reaches $n$, as opposite,
the only change happens when the fourth cell is updated, and it decreases the length.
So, the random walk cannot indefinitely stay in state $n$.

\parpic[r]{$
  \begin{matrix}
    \cdots\ 0\ 1\ 0\ 1\ 0\ 0\ 1\ 0\ \cdots\\[-0.5ex]
    \cdots\ 0\ 1\ 0\ 1\ 1\ 0\ 1\ 0\ \cdots
  \end{matrix}
$}
On the other hand, when the length reaches $1$, one possibility is the
opposite, where updating the fifth cell leads to coalescence.

\parpic[r]{$
  \begin{matrix}
    \cdots\ 0\ 1\ 0\ 1\ 1\ 0\ 1\ 0\ \cdots\\[-0.5ex]
    \cdots\ 0\ 1\ 0\ 0\ 1\ 0\ 1\ 0\ \cdots
  \end{matrix}
$}
The other possibility is opposite, where updating the fifth then the fourth cell
leads to the former possibility.
In each case, coalescence happens with a constant non zero probability.
One step of this random walk takes an expected time $O(n)$, the total expected
time of the one word step is thus $O(n^3)$
(details on expected time can be found in~\cite{async_dqeca_formal}).

So rule \rn6 is coalescing.
The only difference of \rn7 is that $000$ leads to $010$, which does not
affect the proof (only the increasing phase is easier). \qed
\end{proof}
\begin{remark}
If $n$ is even, the proof is valid
until there is only one word, at which point
we get a configuration
without $00$ nor $11$. There are two such configurations, if both copies have
the same, it is coalescence, otherwise both copies perfectly disagree (definitively).
Both happen experimentally.
\end{remark}

\begin{theorem}
\label{main-theo}
For the fully asynchronous dynamic,
there are non trivial state-minimal coalescing cellular automata with an arbitrarily large number of
states, and therefore infinitely many non trivial state-minimal coalescing CA.
\end{theorem}
\begin{proof}
Let ${\cal A}^2$ be the product of a CA
${\cal A}=(Q,\,d,\,V,\,\delta,\,n,\,\mu)$ by
itself, defined as $(Q^2,\,d,\,V,\,\delta^2,\,n,\,\mu)$
where $\delta^2\big((a,b),\,(c,d),\,(e,f)\big):= \big(\delta(a,c,e),\,\delta(b,d,f)\big)$.
Intuitively, ${\cal A}^2$ is the automaton we get by superposing
two configurations of ${\cal A}$ and letting both evolve according to $\delta$,
but with the same $M_t$. If ${\cal A}$ is state-minimal, so is ${\cal A}^2$.

Let ${\cal A}$ be a coalescing CA. Then ${\cal A}^2$ converges in
polynomial expected time towards a configuration of states all in
$\set{(q,q)}{q\in Q}$.
From this point, ${\cal A}^2$ simulates ${\cal A}$ (by a mere projection of $Q^2$
to $Q$) and is therefore coalescing (with an expected time at most
twice as long); and so are $({\cal A}^2)^2$, $\big(({\cal A}^2)^2\big)^2$, etc.
We have built an infinite sequence of CA with increasing size.\qed
\end{proof}

\begin{proposition}
\label{15-170-semi-coalescent}
  \rn{15} and \rn{170}, for both asynchronous dynamics, coalesce
  or end in total disagreement, each case with probability $1/2$.
\end{proposition}
\begin{proof}
  \rn{170} (shift) means ``copy your right neighbor''.
  The configurations agree on a cell
  if and only if
  they agreed on the right neighbor
  before this cell was updated.
  So, it is a CA with two states: agree or disagree.
  This CA is still \rn{170}.
  This rule converges in polynomial time towards $0^*$
  (corresponding to coalescence) or $1^*$ (full disagreement) \cite{async_dqeca_formal}.
  By symmetry, each case has probability $1/2$.

  \rn{15} means ``take the state opposed to the one of your right neighbor'',
  and the proof is identical (the quotient CA is still \rn{170}). \qed
\end{proof}

\section{Experimental Study and Phase Transition}
\label{expérimental-study}
In this section, we describe experimental results in the context of partially
asynchronous dynamic. We show that many ECA exhibit coalescence and make a finer
classification.
Specifically, we observe that some ECA undergo a phase transition for this property when
$\alpha$ varies. We experimentally show that this phase
transition belongs to the universality class of directed percolation.

\subsection{Classifying CA with Respect to Coalescence}
\paragraph{Protocol}
We call \emph{run} the temporal evolution of a CA when all parameters
(rule, size, $\alpha$ and an initial configuration) are chosen.
We stop the run when the CA has coalesced, or when a predefined maximum running
time has been reached.

Let us describe the parameters we used.
We set $n=500$ and $n=2\,000$ and got the same results.
%
\cite{async_dqeca_formal} showed rigorously that $\alpha$ close to $1$ (more
updates) does not mean faster convergence (indeed, it is proportional to $\frac{1}{\alpha(1-\alpha)}$).
We thus repeat each run three times: for $\alpha\in\{0.05, 0.50, 0.95\}$.
The maximum number of time steps is equal to a few times $n^2$.
For each rule, we do 30 runs: 10 random initial configurations (to ensure
coherence) times 3 values of $\alpha$.

\paragraph{Results}
We get the following empirical classes of behaviour:
\begin{itemize}
\item[a/] Some CA never coalesce (or take a too long time to be observed):
  \rn{4}, \rn{5}, \rn{12}, \rn{13}, \rn{25},
  \rn{28}, \rn{29}, \rn{33}, \rn{36}, \rn{37}, \rn{41}, \rn{44}, \rn{45},
  \rn{51}, \rn{54}, \rn{60}, \rn{72}, \rn{73}, \rn{76}, \rn{77}, \rn{78},
  \rn{90}, \rn{94}, \rn{104} \rn{105}, \rn{108}, \rn{122}, \rn{132}, \rn{140},
  \rn{142}, \rn{150}, \rn{156}, \rn{164}, \rn{172}, \rn{200}, \rn{204},
  \rn{232}.

\item Some CA coalesce rapidly.
  \begin{itemize}
  \item[b/] The trivial way to do this is to converge to a unique fixed point. One can
    consider the two copies independently, and wait for them to reach the fixed point,
    the CA has then coalesced. This is the case for
    \rn{0}, \rn{2}, \rn{8}, \rn{10}, \rn{24}, \rn{32}, \rn{34}, \rn{38},
    \rn{40}, \rn{42}, \rn{56}, \rn{74}, \rn{128}, \rn{130}, \rn{136}, \rn{138},
    \rn{152}, \rn{160}, \rn{162}, \rn{168}.
  \item[c/] The non trivial rules are
    \rn{3}, \rn{11}, \rn{19}, \rn{35}, \rn{46}, \rn{154}.
  \end{itemize}

\item[d/] Some CA combine two of the three previous behaviors, depending on $\alpha$.
\rn{18}, \rn{26}, \rn{106}, \rn{146} combine
a/ and b/;
\rn{50}, \rn{58}, \rn{134}
combine b/ and c/;
\rn{1}, \rn{9}, \rn{27}, \rn{57}, \rn{62}, \rn{110} and \rn{126}
combine a/ and c/
(see fig.~\ref{fig:time-space}).

\item[e/] Some CA end in either full agreement between configurations (coalescence)
  or full disagreement, depending on the outcome of $M_t$ and the initial configuration:
  \rn{14}, \rn{15}, \rn{23}, \rn{43}, \rn{170}, \rn{178}, \rn{184}.
\item[f/] \rn{6} combines the previous point (for small $\alpha$) with
  b/,
  \rn7 do the same but with
  c/,
  \rn{22} and \rn{30} combine it (for small $\alpha$) with
  a/.

 \end{itemize}
Let us to study 
the phase transition (coalescence or not) when $\alpha$ changes, that is, rules
\rn{1}, \rn{9}, \rn{27}, \rn{57}, \rn{62}, \rn{110} and \rn{126}.
We test the hypothesis of directed percolation.
Some rules (\rn9, \rn{110}, \rn126) show two phase transitions,
one for low $\alpha$, noted \rnnl, one for
high $\alpha$, noted \rnnh.
The ones for low $\alpha$ (\rnnl, \rncdl and \rncvsl) are ``reversed'', that is, coalescence (sub-critical regime)
occurs for higher $\alpha$. \rn{57} is also reversed.

\subsection{Directed Percolation}
\label{directed-perco}
Due to lack of space, we refer to~\cite{hinrichsen-2000-49} for a presentation
of directed percolation, which also explains
\emph{damage spreading}, another point of view on this phenomenon.

\begin{figure}[tbp]
  \centerline{\hbox{
\ifpdf
    \includegraphics[width=0.49\linewidth]{figures/rn110_ur47_step500.png}
    \includegraphics[width=0.49\linewidth]{figures/rn110_ur65_step500.png}
\else
    \psfig{figure=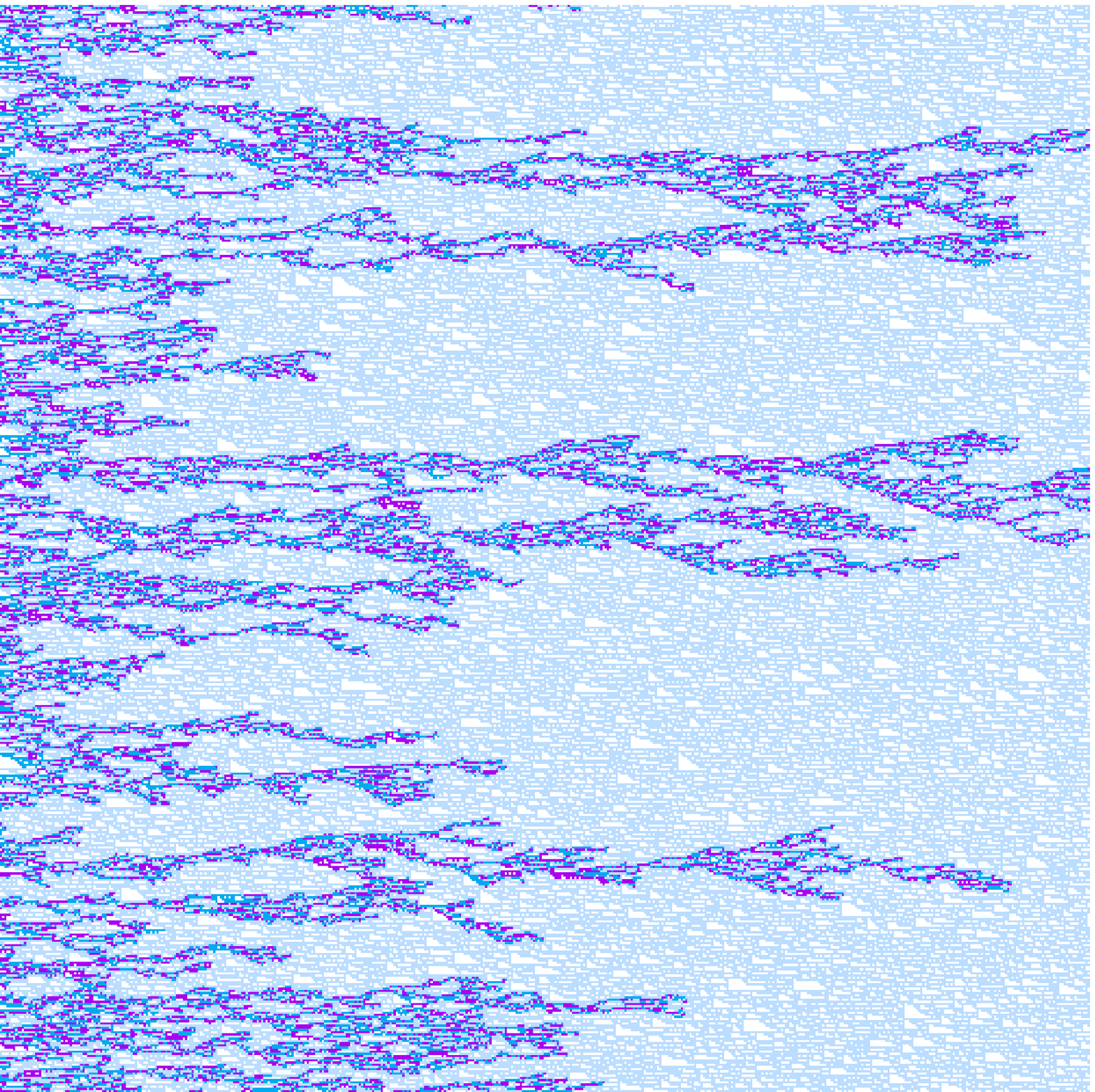,width=0.49\linewidth}
    \psfig{figure=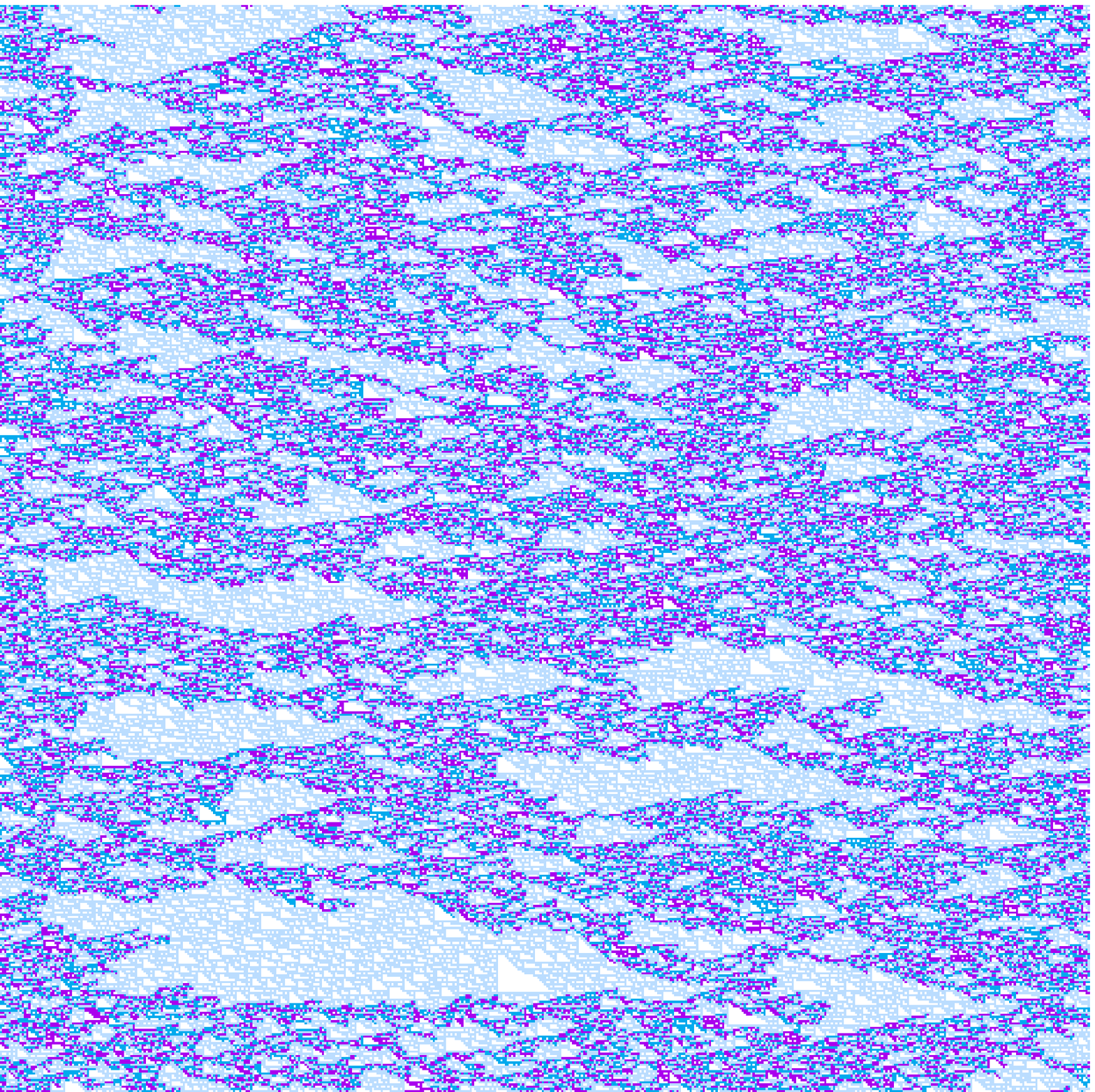,width=0.49\linewidth}
\fi
}}
  \caption{(color online)
    rule \rn{110}, $n=500$.
    Time goes from left to right, during $500$ steps.
    Active sites are dark, coalesced site are light
    (with light blue standing for state $1$, white for $0$).
Left: sub-critical phase ($\alpha=0.47<\alpha_c\simeq 0.566$), branches die.
Right: supercritical phase ($\alpha=0.65>\alpha_c$), active sites spread
}
  \label{fig:time-space}
\end{figure}

Our active sites are the cells where the configurations disagree (density of
such sites is written $\rho$). Percolation appears when varying $\alpha$, see fig.~\ref{fig:time-space}.
The aim is thus to identify $\beta$ assuming that
$\rho(\alpha) = c\,(\alpha-\alpha_c)^{\beta}$ for some $c$ and $\alpha_c$.

\subsubsection{Measure of \texorpdfstring{$\alpha_c$}{critical alpha}}
We use the method described in~\cite{hinrichsen-2000-49}: plot
the density $\rho$ of active sites versus time in logarithmic scale and find the
$\alpha$ value for which one gets a straight line (for $\alpha<\alpha_c$, the AC
coalesce faster, for $\alpha>\alpha_c$, it has a positive asymptotic $\rho$).
We used random initial configuration with each state equiprobable.
To get readable plots we needed up to $n=10^6$ cells and $10^7$ time steps.
We get (recall that $\alpha_c$ is not universal, it is just used to compute
$\beta$):
\\\centerline{
  \begin{tabular}{|l|c@{\ \ }c@{\ \ }c@{\ \ }c@{\ \ }c@{\ \ }c@{\ \ }c@{\ \ }c@{\ \ }c@{\ \ }c|}
    \hline
    rule              &\rn{1} &\rnnl  & \rnnh &\rn{27}&\rn{62}&\rncdl&\rncdh &\rncvsl&\rncvsh&\rn{57}\\
    \hline
    $\alpha_c >\dots$ & 0.102 & 0.073 & 0.757 & 0.856 & 0.598 & 0.073 & 0.566 & 0.101 & 0.720  & 0.749\\
    $\alpha_c <\dots$ & 0.103 & 0.074 & 0.758 & 0.858 & 0.599 & 0.075 & 0.567 & 0.102 & 0.721  & 0.750\\
    \hline
  \end{tabular}
}
Note that the $\alpha_c$ of \rn{1} and \rncvsl, like \rnnl and \rncdl, are very close.

\subsubsection{Measure of \texorpdfstring{$\beta$}{beta}}
\begin{figure}[tbp]
  \centerline{\hbox{
\ifpdf
\includegraphics[width=0.8\linewidth]{figures/fit110.png}
\else
\psfig{figure=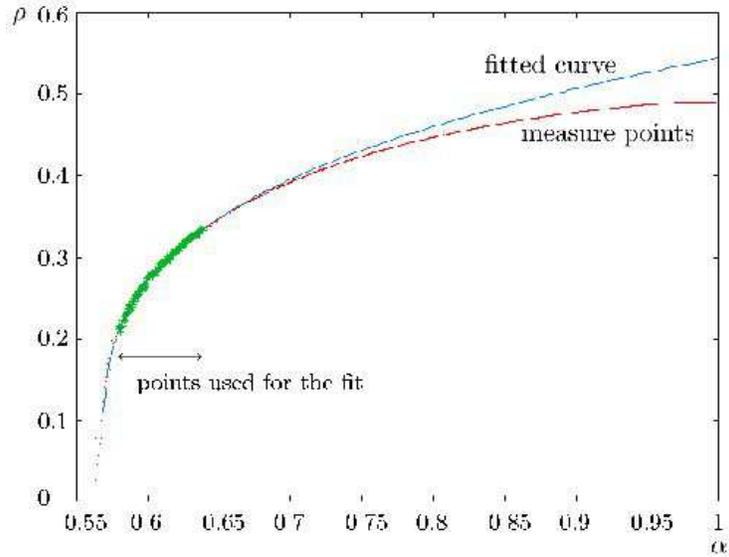,width=0.8\linewidth}
\fi
}}
  \caption{
    $\rho(\alpha)$. Rule \rn{110}, $n=10\,000$.
    Fitted curve
    agrees with points used for the fit (bold).
}
  \label{fig:fit_beta}
\end{figure}

We now plot $\rho$ vs $\alpha$ (fig.~\ref{fig:fit_beta}).
The assumption
$\rho=c\,(\alpha-\alpha_c)^\beta$ is valid only near $\alpha_c$.
To determine which points should be taken into account, we plot $\rho(\alpha)$ on
a logarithmic scale with $x$-origin roughly equal to $\alpha_c$ (precision
does not affect the result).
We keep only the beginning of the curve which is a straight line.
We varied the number of points taken into account to estimate the loss of
precision due to this choice.

\paragraph{Protocol}
\label{sec:protocole_beta}
$n=10\,000$. We let the system evolve for $\Ttr=100\,000$ steps,
then measure $\rho$ during $\Tech=10\,000$ step and compute the average.
We repeat such a run for each $\alpha$ value with a fine sampling.
The fact that the curve is smooth (except very near $\alpha_c$) tells us that
measures do not depend much on the randomness (nor on $\Tech$) and that we do not need error bars.
We also checked that the results do not vary when we change $n$ and $\Ttr$.

The fit gives the following ranges, taking into account uncertainty about
$\alpha_c$ and which points to keep for the fit.
Experimental value for $\beta$ measured on other systems is $0.276$.
\\\centerline{
  \begin{tabular}{|l|c@{\ \ }c@{\ \ }c@{\ \ }c@{\ \ }c@{\ \ }c@{\ \ }c@{\ \ }c@{\ \ }c@{\ \ }c|}
    \hline
    rule           &\rn{1} & \rnnl & \rnnh &\rn{27}&\rn{62}&\rncdl&\rncdh&\rncvsl&\rncvsh&\rn{57}\\
    \hline
    $\beta >\dots$ & 0.265 & 0.270 & 0.273 & 0.258 & 0.270 & 0.270 & 0.271 & 0.250 & 0.260 & 0.248\\
    $\beta <\dots$ & 0.279 & 0.295 & 0.283 & 0.305 & 0.281 & 0.291 & 0.281 & 0.276 & 0.276 & 0.281\\
    \hline
  \end{tabular}
}
As expected, this model seems to belong to the universality class of directed
percolation (except perhaps \rn{27}, due to higher noise and thus lack of precision).

Source code is available on \verb+cimula.sf.net+.



\bibliographystyle{splncs}
\bibliography{biblio}

\end{document}